**Electron-Phonon Dephasing in Ultrathin Disordered Films: From Power Laws to Diagnostic Maps**

**Eugeniy Yu. Beliayev**

B. Verkin Institute for Low Temperature Physics and Engineering, Kharkiv, Ukraine

## Abstract

Electron-phonon dephasing in ultrathin disordered metallic films is commonly summarized by a power law, $\tau_{e-ph}^{-1} = AT^{p}$. However, experiments on thin metals, alloys, and superconducting films report effective exponents close to 2, 3, 4, and intermediate nonuniversal values. We argue that in ultrathin supported films the exponent should not be treated as a universal material constant, but as a crossover observable. Its interpretation requires at least three physical coordinates: the clean-to-dirty parameter $q_T l$, the phonon confinement and film-substrate acoustic coupling axis, and the microscopic character of disorder, ranging between the limiting cases of fully dragged and partially static scattering potentials.

This diagnostic view is grounded in established theoretical limits. Fully dragged disorder potentials lead to dirty-limit suppression of electron-phonon relaxation, $\tau_{e-ph}^{-1} \propto T^{4} \times l$ [1], whereas static disorder components can enhance the effective interaction and produce $\tau_{e-ph}^{-1} \propto T^{2}/l$ [2]. Electron-phonon-impurity interference and transverse phonons provide additional routes to $T^{2}$-type behavior in thin films. The same effective exponent can therefore have different microscopic origins and must be interpreted together with the prefactor, disorder dependence, film thickness, and acoustic boundary condition [2] [3].

We demonstrate this approach using $Ar^{+}$-ion-irradiated Au films as a controlled-disorder series. In these films, the fitted electron-phonon exponent remains near $p \sim 2$ for low values of the pure–dirty crossover temperature $T_{\mathrm{tr}} \simeq \hbar s/(k_B l)$, and then increases toward $p \sim 2.8$ as $T_{\mathrm{tr}}$ rises with disorder. At the same time, the prefactor trend indicates suppression of electron-phonon scattering with decreasing mean free path, consistent with a dirty-limit tendency rather than static-disorder enhancement. The fact that $p$ remains well below the bulk dirty-limit benchmark $p = 4$ points to the continuing role of thin-film or film/substrate phonon effects. We propose a diagnostic map that organizes these regimes and provides a practical framework for comparing electron-phonon dephasing data in ultrathin disordered films [4] [5].

## 1. Introduction: why one exponent is not enough

Electron-phonon dephasing in ultrathin disordered metallic films is often reduced to a single question: what is the exponent $p$ in $\tau_{e-ph}^{-1} = A \cdot T^{p}$ ? This question is useful, but it is also misleading. The literature contains robust examples of $p$ close to 2, 3, 4, and many intermediate effective exponents. Such diversity is sometimes treated as experimental scatter or as failure to reach a clean asymptotic regime. In ultrathin films, however, this diversity may itself be a physical signal.

The fitted exponent is a local summary of several coupled crossovers. First, the electronic motion may cross from clean-like to dirty-like behavior relative to the thermal phonon wavelength. Second, the phonon system may cross from bulk-like behavior to confined, guided, or substrate-modified modes. Third, the disorder potential itself may change character: two samples with similar elastic mean free path may contain very different scatterers, such as impurities, vacancies, boundaries, surface disorder, irradiation-induced defects, pores, or interface damage.

The clean-to-dirty crossover is usually expressed through $q_T l$, where $q_T$ is the thermal phonon wave vector and $l$ is the elastic mean free path. When $q_T l > 1$, electrons move ballistically over the phonon wavelength. When $q_T l < 1$, electron motion is diffusive on that scale, and the electron-phonon vertex is modified by disorder. But $q_T l$ is not sufficient in ultrathin supported films. The ratio between phonon wavelength and film thickness, and the degree of acoustic coupling to the substrate, also enter the problem [2] [6] [7].

The aim of this paper is to formulate a diagnostic map for interpreting effective electron-phonon exponents in ultrathin disordered films. The map does not replace microscopic theory. Rather, it organizes known theoretical limits and experimental observations into a framework that helps answer a more useful question: what does the measured exponent and prefactor reveal about disorder and film phonons?

## 2. Effective exponent as a crossover representation

For a true power law, the exponent $p$ is constant. In crossover regimes, a more informative representation is the local logarithmic exponent

$$p_{eff}(T) = \frac{d \ln \tau_{e-ph}^{-1}}{d \ln T}. \tag{1}$$

This quantity is not new physics by itself. It is simply the local slope of the relaxation rate on a log-log plot. Its usefulness lies in preventing overinterpretation of a single global fit. If $\tau_{e-ph}^{-1}$ follows a pure power law, $p_{\text{eff}}$ is constant. If the system crosses from one regime to another, $p_{\text{eff}}$ traces that crossover.

The important point is that $p_{\text{eff}}$ cannot be interpreted alone. In ultrathin films it acquires physical meaning only when placed together with the relevant regime parameters: $q_T l$, $q_T d$, $\Gamma_{fs}$, and $\eta_{\text{drag}}$. Here $q_T l$ describes the clean-to-dirty electronic crossover, $d$ is the film thickness, and $q_T d$ measures whether thermal phonons are bulk-like or affected by confinement. The parameter $\Gamma_{\text{fs}}$ denotes the effective film–substrate acoustic coupling, while $\eta_{\text{drag}}$ is a phenomenological label for the limiting character of disorder, ranging from fully dragged scattering potentials to static or incompletely dragged components.

This formulation is intended as a diagnostic framework rather than a microscopic classification scheme. We do not claim that $p_{\text{eff}}$ uniquely identifies a microscopic mechanism. Instead, it is used as a compact diagnostic representation. The same apparent $T^2$ behavior, for example, may arise from static disorder [2], transverse-phonon interference [3] [8], or film/substrate acoustic effects [7] [9]. In a given disorder series, the key discriminator is the combined behavior of $p$ and the prefactor $A(l)$. Independent variations of thickness and substrate are then needed to separate electronic disorder effects from phonon-confinement and film–substrate acoustic effects.

## 3. Schematic diagnostic map

Fig. 1 schematically summarizes the diagnostic map. The horizontal axis represents $q_T l$, separating clean-like and dirty-like electron motion relative to the thermal phonon wavelength. The vertical axis represents phonon confinement and film-substrate acoustic coupling. The dashed guides $q_T l = 1$ and $q_T d \sim 1$ are crossover markers, not sharp phase boundaries. The third coordinate is the limiting character of disorder: fully dragged scattering potentials on one side and partially static disorder on the other.

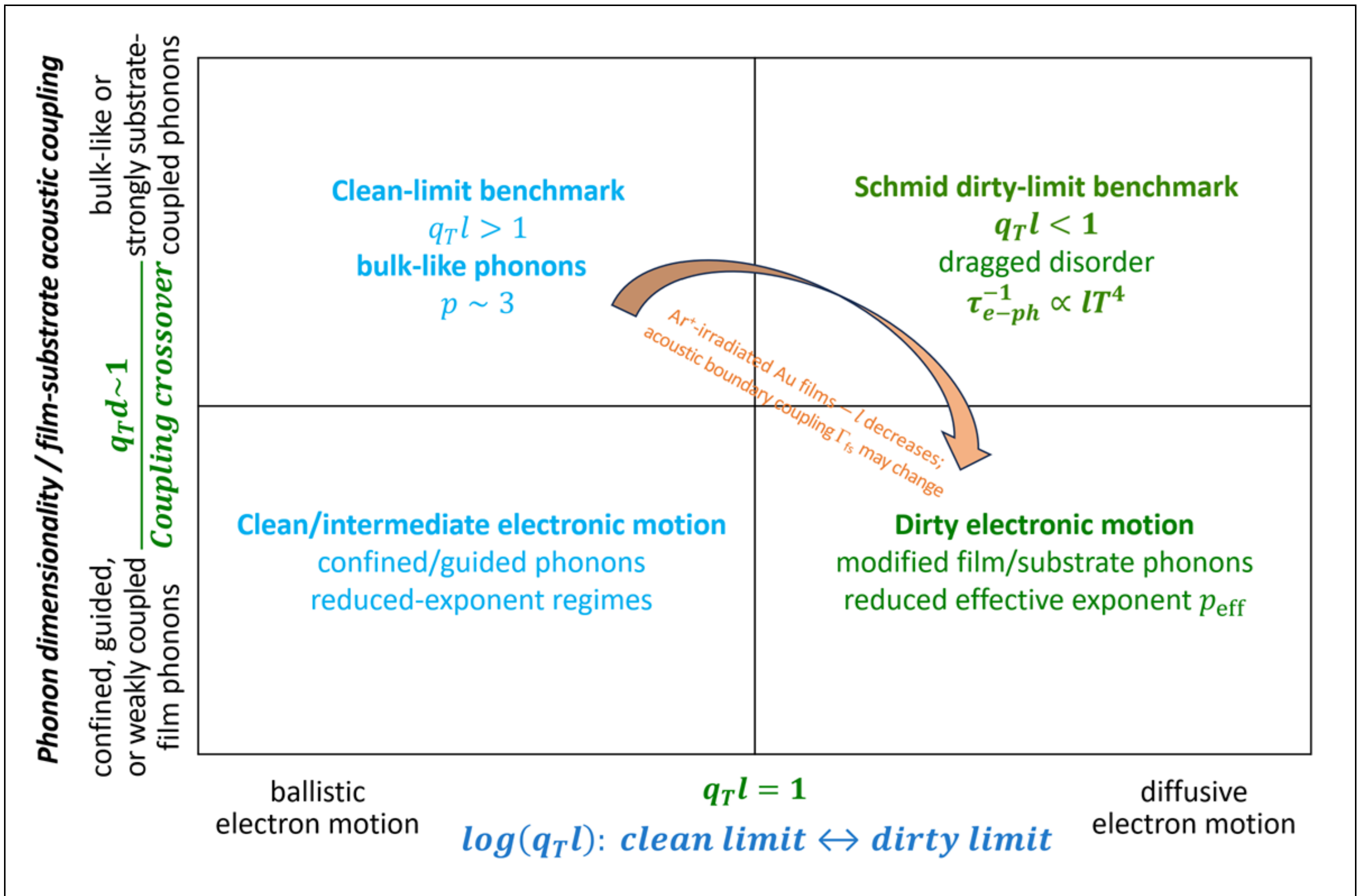

Fig. 1. Diagnostic map for electron–phonon dephasing in ultrathin films. The horizontal axis represents $q_T l$, while the vertical axis represents phonon confinement and film-substrate acoustic coupling. The arrow indicates that $Ar^+$-ion irradiation may simultaneously reduce the electronic mean free path $l$ and modify the acoustic boundary condition $\Gamma_{fs}$.

The key point is that an external perturbation may move a film along more than one axis. Ion irradiation of an ultrathin supported film is a particularly important example. It decreases $l$, pushing the system toward the dirty side, but it may also produce holes, local stress relaxation, defect complexes, and interface damage. Such changes may modify the acoustic boundary condition and therefore the phonon spectrum relevant for electron-phonon relaxation [4] [5].

Thus, an irradiated ultrathin film should not automatically be interpreted as a one-parameter $q_T l$ experiment. It may be a coupled perturbation in both electronic disorder and film/substrate phonons.

## 4. Theoretical benchmarks

A diagnostic map is useful only if its limiting regions are tied to known theoretical benchmarks.

The first dirty-limit benchmark is the Schmid result. Schmid showed that in an impure metal the effective electron–phonon interaction is reduced when the phonon wavelength becomes longer than the electronic mean free path, and that the inelastic collision time with phonons increases as impurities are added [1]. In modern notation this corresponds to dirty-limit suppression in the $q_T l < 1$ regime, commonly written as $\tau_{e-ph}^{-1} \propto T^4 l$. Sergeev and Mitin later recovered this dragged-disorder limit and showed that, if the disorder potential contains a static or incompletely dragged component, interference between scattering processes can enhance the effective electron–phonon interaction and yield $\tau_{e-ph}^{-1} \propto T^2/l$ [2].

These laws should be treated as limiting benchmarks rather than as a binary classification of real defects. Real impurities have mass, real boundaries may be partially compliant, and irradiation-induced defects may include several kinds of scattering centers.

A second route to $T^2$-like behavior involves electron-phonon-impurity interference and transverse phonons. In this case, $p$ close to 2 does not necessarily indicate static disorder enhancement. It may instead reflect a transverse-phonon channel or an intermediate thin-film regime [8] [3].

A third route comes from the phonon side. In ultrathin films, the thermal phonon wavelength may become comparable to the film thickness. In supported films, the substrate is not passive: acoustic mismatch, guided modes, phonon escape, and weak or strong adhesion can all modify the effective phonon spectrum. Thus, film/substrate phonons form an independent axis of the diagnostic map [7] [9] [10].

## 5. Controlled-disorder series: Ar-ion-irradiated Au films

$Ar^+$-ion-irradiated Au films provide a useful controlled-disorder series because the disorder was modified in an approximately fixed ultrathin metallic geometry. The films were about 10 nm thick, and $Ar^+$-ion irradiation produced a series of samples with sheet resistances ranging from a few ohms to several hundred ohms [4] [5]. The electron–phonon contribution was represented as

$$\tau_{e-ph}^{-1} = A_p(l)T^p.$$

The original analysis showed that $p \simeq 2$ in low-resistance films and increases with stronger disorder, while the prefactor $A_p(l)$ indicates a decrease of the electron-phonon interaction rate as $l$ decreases [4] [5].

The crossover temperature $T_{\mathrm{tr}}$ is defined by

$$q_T l \sim 1, \quad T_{\mathrm{tr}} \simeq \frac{\hbar s}{k_B l},$$

where $s$is the relevant sound velocity (longitudinal or transverse, depending on the dominant electron-phonon coupling). Thus, $T_{\mathrm{tr}}$ increases as the electron mean free path decreases. In the Au dataset [5], $T_{\mathrm{tr}}$ rises from a few kelvins in weakly disordered films to more than 100 K in the most disordered sample. The reorganized data therefore provide a direct controlled-disorder series in the ($T_{\mathrm{tr}}$, $p$) plane. This controlled-disorder series is summarized in Fig. 2, where the fitted exponent $p$ is replotted as a function of the pure–dirty crossover temperature $T_{tr}$, emphasizing the crossover character of the fitted exponent.

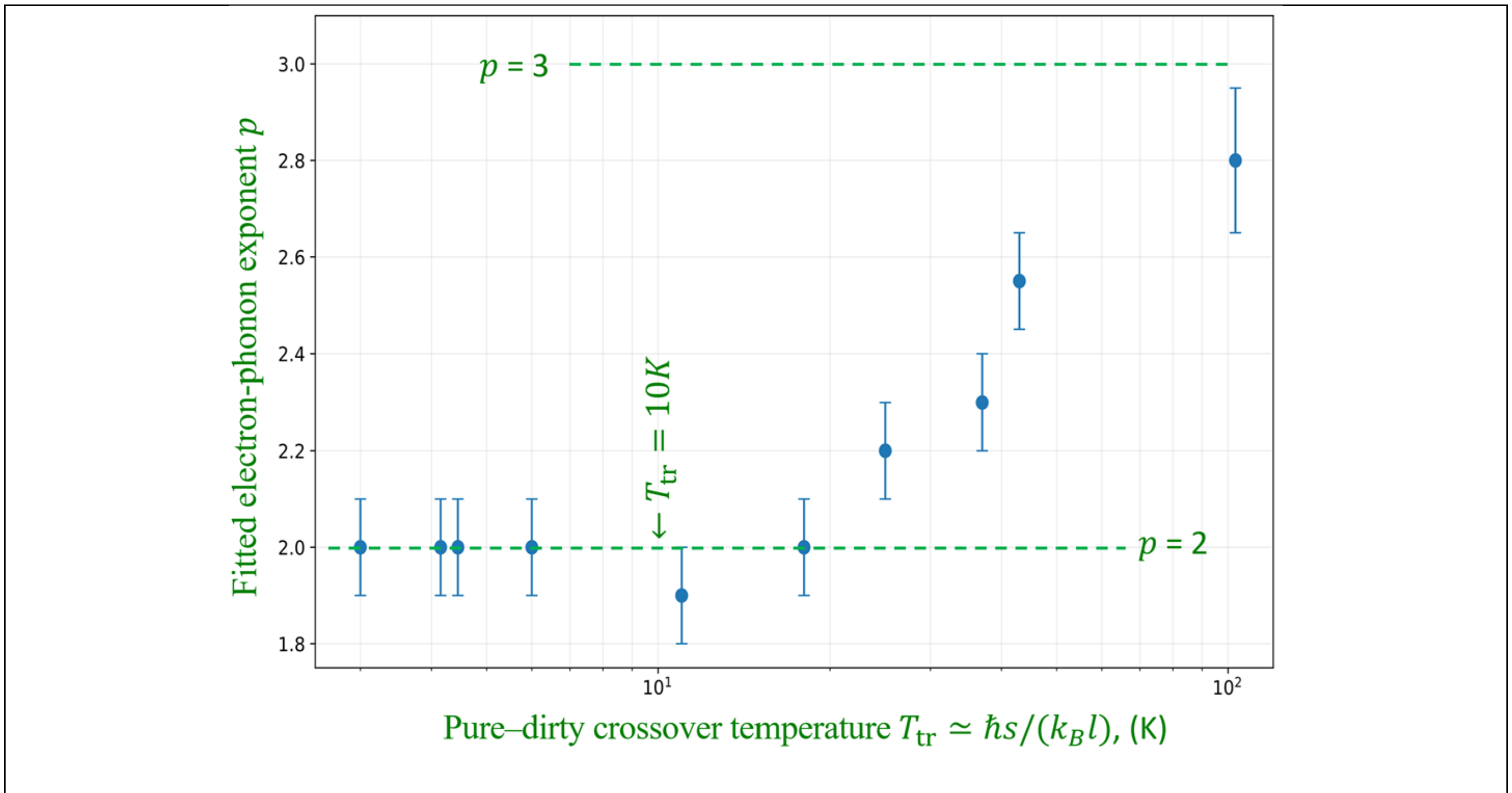

Fig. 2. Ar+-ion-irradiated Au films as a controlled-disorder series. The fitted electron–phonon exponent $p$ is replotted as a function of the pure–dirty crossover temperature $T_{\rm tr} \simeq \hbar s/(k_B l)$. Data points are digitized from Ref. [5]; the lines are guides to the eye. The exponent remains close to 2 for $T_{\rm tr} \lesssim 10$ K and increases toward 2.8 as $T_{\rm tr}$ enters and exceeds the experimental temperature window. This representation emphasizes the crossover character of the fitted exponent rather than treating $p$ as a universal constant.

The important observation is not simply that $p$ close to 2 appears. The significant point is the combined behavior of exponent and prefactor. As disorder increases, $p$ rises from approximately 2 toward 2.8, but the electron-phonon scattering rate is suppressed rather than enhanced. In the original analysis of the 20 K data, one group of samples showed weak dependence on $l$, while the more strongly disordered branch was described approximately by $\tau_{ep}^{-1} \propto l$ [5].

This means that the Ar+-irradiated Au series is not a static-disorder $T^2/l$ case. It is better understood as a dirty-limit suppression trend occurring in a film where low-dimensional or film/substrate phonon effects keep the effective exponent below the bulk dirty-limit benchmark $p = 4$.

**Table 1. Ar+-ion-irradiated Au films as a controlled-disorder series.**

| $R_\square$ (ohm) | $T_{\rm tr}$ (K) | $p$ | Prefactor/rate trend | Diagnostic implication |
|---|---|---|---|---|
| 10.6 | 3.0 | 2.0 +/- 0.1 | weak $l$-dependence | low-disorder $T^2$-like regime |
| 14.6 | 4.3 | 2.0 +/- 0.1 | weak $l$-dependence | low-disorder $T^2$-like regime |
| 14.7 | 4.3 | 2.0 +/- 0.1 | weak $l$-dependence | low-disorder $T^2$-like regime |
| 20.4 | 6.0 | 2.0 +/- 0.1 | weak $l$-dependence | low-disorder $T^2$-like regime |
| 44.4 | 11 | 1.9 +/- 0.1 | crossover region | boundary region |
| 61.4 | 18 | 2.0 +/- 0.1 | onset of stronger $l$-dependence | crossover onset |
| 85 | 25 | 2.2 +/- 0.1 | $A_p$ decreases with disorder | dirty influence begins |
| 143 | 37 | 2.3 +/- 0.1 | decreasing $A_p$ | dirty influence |
| 161 | 43 | 2.55 +/- 0.1 | suppressed *e-ph* rate | stronger dirty influence |
| 437 | 103 | 2.8 +/- 0.15 | $\tau_{ep}^{-1} \propto l$ trend | strongest disorder in dataset |

$T_{tr}$ is the approximate temperature at which $q_T l \sim 1$. The prefactor column summarizes the behavior discussed in the original Au-film analysis: the low-disorder group shows weak dependence on $l$, whereas the more disordered group shows a decrease of the electron-phonon scattering rate with decreasing $l$, consistent with dirty-limit suppression rather than static-disorder enhancement [5].

## 6. Prefactor trend and differential diagnosis

The prefactor trend is essential because the exponent alone does not diagnose the mechanism. To make this point explicit, approximate coordinates were extracted from the published $\tau_{e-ph}(20\text{ K})$ versus $l$ graph and replotted as Fig. 3.

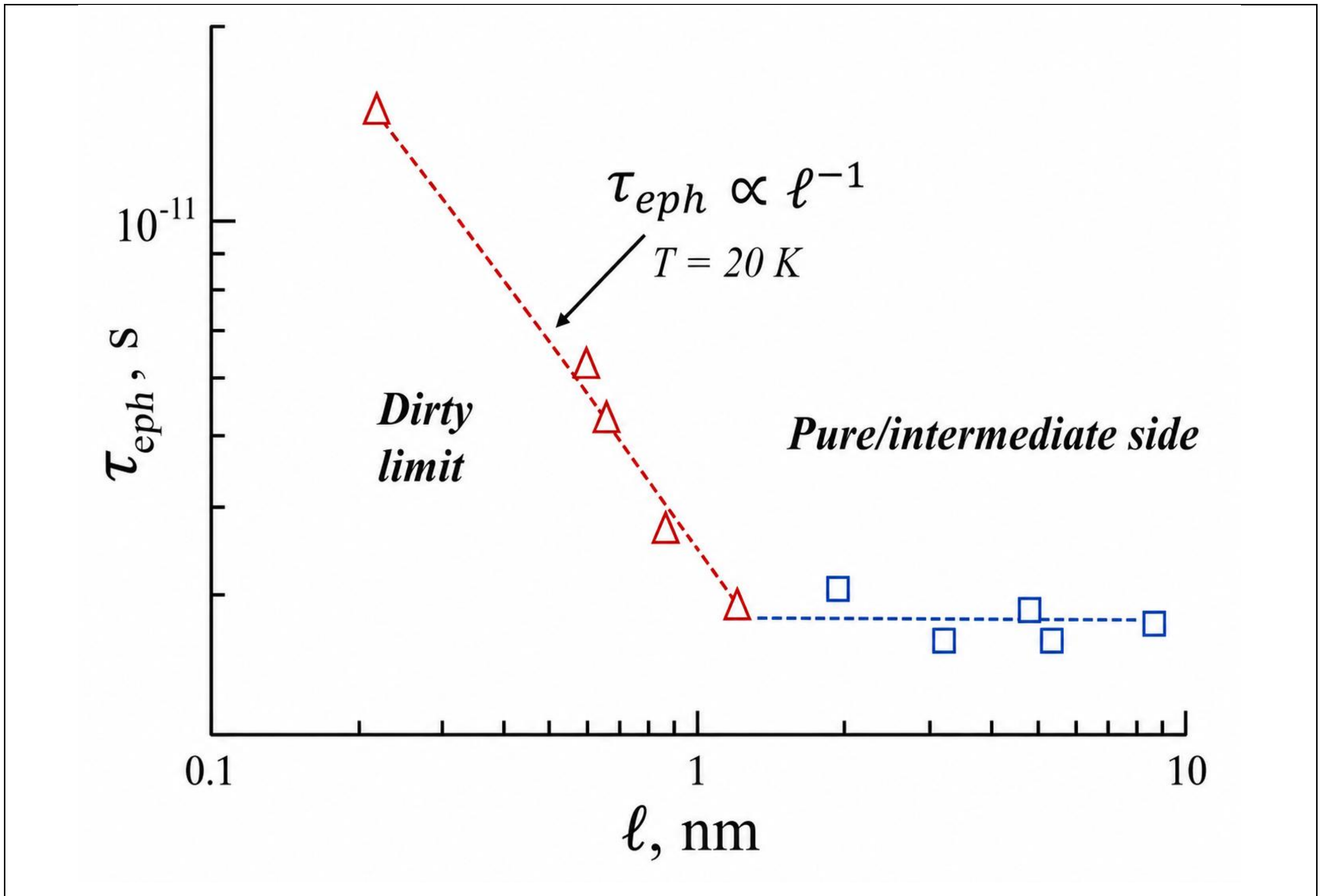

Fig. 3. Approximate reconstruction of $\tau_{e-ph}(20\ K)$ versus mean free path $l$ in $Ar^+$-ion-irradiated Au films. The reconstruction is based on digitized coordinates from the published figure [5] and is intended as an illustrative replot, not as primary data. The short-$l$ branch shows a pronounced increase of $\tau_{e-ph}$ as $l$ decreases, corresponding to suppression of $\tau_{e-ph}^{-1}$. The large-$l$ branch shows much weaker $l$-dependence. The qualitative interpretation follows the original analysis, which identified one group with weak $l$-dependence and another group with approximately $\tau_{ep}^{-1} \propto l$.

This replot reinforces the differential diagnosis. In a static-disorder-enhanced $T^2/l$ scenario, the electron-phonon scattering rate would grow as $l$ decreases. In the Au data, the opposite trend is observed on the disorder-dependent branch. Therefore, the $T^2$-like behavior at low disorder should not be interpreted as static-potential enhancement. Instead, the Au films appear to follow a mixed trajectory: dirty-limit suppression develops as $l$ decreases, while thin-film or film/substrate phonons keep the effective exponent reduced [5].

The same point can be stated more generally. A $T^2$-like dependence is not unique and should not be assigned to a microscopic mechanism by exponent alone. In a given disorder series, the key discriminator is the combined behavior of $p$ and the prefactor $A(l)$. Independent variations of thickness and substrate coupling provide additional tests for separating electronic disorder effects from phonon-confinement and film/substrate acoustic effects. Table 2 summarizes this differential diagnosis.

**Table 2. Differential diagnosis of $T^2$-like electron–phonon relaxation in ultrathin disordered films.**

| Observation | Likely interpretation | Required check |
|---|---|---|
| $p \sim 2$, rate increases with disorder, $A$ proportional to $1/l$ | static or incomplete-drag disorder contribution | vary disorder at fixed $d$ and substrate |
| $p \sim 2$, weak $A(l)$ dependence | transverse phonon / electron-phonon-impurity interference | compare with $T^2$ resistivity contribution |
| $p \sim 2$, strong $d$ or substrate dependence | film/substrate phonon spectrum | vary thickness or acoustic boundary |
| $p \sim 2 \rightarrow 2.8$, but rate decreases as $l$ decreases | dirty-limit suppression mixed with film phonons | $Ar^+$-irradiated Au series |

The table is intended to prevent overinterpretation of the exponent alone. In the $Ar^+$-irradiated Au series, the decisive feature is not $p \simeq 2$ at low disorder, but the combined evolution of $p$ and the prefactor: $p$increases toward 2.8 while the electron–phonon scattering rate decreases with decreasing $l$. This places the Au series in the dirty-suppression-plus-film-phonon sector rather than in a static-disorder-enhanced $T^2/l$ scenario.

## 7. Bi films as a thermal route to the dirty limit

Disordered Bi films provide a complementary calibration case. In these films, dirty-limit weakening of electron-phonon relaxation was observed not only by changing disorder but also by lowering temperature. This matters because $q_T l$ is temperature dependent: a fixed sample can move into the dirty limit as the thermal phonon wavelength grows on cooling. Thus, the dirty condition is not merely a label attached to a sample; it is a temperature-dependent coordinate [11].

The Bi experiments therefore support the diagnostic-map view from a different direction. They show that the same material system can move through the map thermally, without changing its structural disorder. This is precisely why a single fitted exponent over a broad temperature interval may obscure the underlying crossover.

## 8. Benchmark limits from later experiments

The map also organizes later benchmark experiments. In disordered Cu and Au films studied by SINIS thermometry at subkelvin temperatures, a $T^4$-type energy-relaxation law was observed and interpreted as disorder-induced weakening of electron-phonon interaction. This provides a low-temperature benchmark for the dirty-limit suppression side of the map [12].

In AuPd wires, electron-phonon dephasing with a $T^2$ law was observed in a higher-temperature range and interpreted as scattering from vibrating defects and impurities. This provides a benchmark for defect-assisted $T^2$-type dephasing [13].

These two results are not contradictory. They occupy different regions of the diagnostic map. The same apparent exponent problem appears because experiments probe different combinations of $q_T l$, film geometry, disorder character, and measured observable.

The same diagnostic issue appears in modern ultrathin superconducting detector films, where electron-phonon scattering, phonon escape, and internal bottlenecks determine device response. NbN and WSi films provide examples in which reduced phonon density of states and internal phonon bottlenecks become central, but these materials are treated here only as boundary cases and motivation rather than as the main subject [10] [14] [15].

## 9. Experimental tests and outlook

A direct test of the diagnostic map would be to prepare two series of ultrathin Au films with comparable l and thickness but different acoustic boundary conditions: strongly substrate-coupled, weakly adhered, and suspended. The map predicts that, at fixed $q_T l$, the effective exponent and crossover width should depend systematically on film-substrate acoustic coupling, while the prefactor scaling with l should remain primarily tied to electronic disorder.

A second test would be to modify disorder by two different methods, such as ion irradiation and low-temperature annealing, at fixed substrate and thickness. If both procedures produce the same $l$ but different $p$ or $A_p(l)$, the difference would indicate that microscopic disorder character, not only mean free path, controls electron-phonon dephasing.

Such tests would separate electronic dirty-limit effects from phonon-boundary effects. They would also clarify whether the reduced exponents observed in ultrathin supported films originate primarily from disorder dynamics, transverse phonons, or film/substrate acoustic modes.

## 10. Conclusion

Electron-phonon dephasing exponents in ultrathin disordered films should not be treated as universal material constants. They are effective crossover observables. Their interpretation requires at least three coordinates: $q_T l$, phonon confinement / film-substrate acoustic coupling, and the limiting character of disorder as dragged or static.

$Ar^+$-ion-irradiated Au films provide a concrete controlled-disorder series. The exponent increases from approximately 2 to 2.8 as the crossover scale $T_{tr}$ rises with disorder, while the prefactor indicates suppression of electron-phonon scattering with decreasing $l$. This combined behavior is consistent with dirty-limit influence, but not with a simple bulk $T^4$ law. The reduced exponent points to thin-film or film/substrate phonon effects [4] [5].

We emphasize that the present Au data do not prove a modification of $\Gamma_{fs}$. The observed saturation of $p$ below 4 can equally be explained by the fact that the experimental temperature window does not reach the deep asymptotic dirty limit, or by a coupled change in $\Gamma_{fs}$. The diagnostic map does not require us to know which effect dominates; it simply brackets the trajectory. The key diagnostic statement is that a one-parameter $q_T l$ interpretation is insufficient to explain the full dataset, regardless of whether the missing axis is populated by $\Gamma_{fs}$ or by finite-temperature crossover width.

The useful question is therefore not which exponent is correct, but what the exponent reveals when placed on the correct diagnostic map. In this sense, electron-phonon dephasing can serve as a compact representation of both microscopic disorder and phonon boundary conditions in ultrathin films.